\documentclass{article}
\usepackage{graphicx}
\usepackage{slashed}
\usepackage{cancel}

\usepackage{amssymb}
\usepackage{amsmath}
\usepackage{mathptmx}       
\usepackage{helvet}         
\usepackage{courier}        
\usepackage{type1cm}        
\usepackage{makeidx}         
\usepackage{graphicx}        
\usepackage{multicol}        
\usepackage[bottom]{footmisc}

\def\d{\mathrm{d}}

\def\bC{{\bf C}}

\def\Im{\mathop{\rm Im}\nolimits}
\def\Re{\mathop{\rm Re}\nolimits}

\def\tg{\mathop{\rm tg}\nolimits}
\def\th{\mathop{\rm th}\nolimits}
\def\ch{\mathop{\rm ch}\nolimits}
\def\sh{\mathop{\rm sh}\nolimits}

\def\wt{\widetilde}

\def\bC{{\bf C}}

\def\Im{\mathop{\rm Im}\nolimits}
\def\Re{\mathop{\rm Re}\nolimits}

\def\tg{\mathop{\rm tg}\nolimits}
\def\th{\mathop{\rm th}\nolimits}
\def\ch{\mathop{\rm ch}\nolimits}
\def\sh{\mathop{\rm sh}\nolimits}

\def\wt{\widetilde}

\def\interior#1{\setbox1=\hbox{$#1$}\rlap{$#1$}\kern0.4\wd1\raise1.1\ht1%
\hbox{$\scriptstyle \circ$}}

\def\boxit#1#2{\setbox1=\hbox{\kern#1{#2}\kern#1}%
\dimen1=\ht1 \advance \dimen1 by #1 \dimen2=\dp1 \advance \dimen2 by #1
\setbox1=\hbox{\vrule height\dimen1 depth\dimen2\box1\vrule}%
\setbox1=\vbox{\hrule\box1\hrule}%
\advance \dimen1 by .4pt \ht1=\dimen1 \advance \dimen2 by .4pt \dp1=\dimen2
\box1\relax}
\def\endprf{\raise .5ex\hbox{\boxit{2pt}{\ }}}

\def\ifundefined#1{\expandafter\ifx\csname#1\endcsname\relax}
\def\ttheta{{\zeta}}
\def\beq{\begin{equation}}
\def\endq{\end{equation}}
\def\eeq{\end{equation}}
\def\beqa{\begin{eqnarray}}
\def\bea{\begin{eqnarray}}
\def\endqa{\end{eqnarray}}
\def\eea{\end{eqnarray}}
\def\x{{\theta}}

\def\P{{\bf P}}
\def\Q{{\bf Q}}

\def\bC{{\bf C}}

\def\Im{\mathop{\rm Im}\nolimits}
\def\Re{\mathop{\rm Re}\nolimits}

\def\tg{\mathop{\rm tg}\nolimits}
\def\th{\mathop{\rm th}\nolimits}
\def\ch{\mathop{\rm ch}\nolimits}
\def\sh{\mathop{\rm sh}\nolimits}

\def\wt{\widetilde}

\def\interior#1{\setbox1=\hbox{$#1$}\rlap{$#1$}\kern0.4\wd1\raise1.1\ht1%
\hbox{$\scriptstyle \circ$}}

\def\boxit#1#2{\setbox1=\hbox{\kern#1{#2}\kern#1}%
\dimen1=\ht1 \advance \dimen1 by #1 \dimen2=\dp1 \advance \dimen2 by #1
\setbox1=\hbox{\vrule height\dimen1 depth\dimen2\box1\vrule}%
\setbox1=\vbox{\hrule\box1\hrule}%
\advance \dimen1 by .4pt \ht1=\dimen1 \advance \dimen2 by .4pt \dp1=\dimen2
\box1\relax}
\def\endprf{\raise .5ex\hbox{\boxit{2pt}{\ }}}

\def\ifundefined#1{\expandafter\ifx\csname#1\endcsname\relax}

\def\beq{\begin{equation}}
\def\endq{\end{equation}}
\def\beqa{\begin{eqnarray}}
\def\endqa{\end{eqnarray}}
\def\twofcn{{F}}

\renewcommand{\cosh}{\ch}\renewcommand{\sinh}{\sh}
\renewcommand{\tanh}{\th}\renewcommand{\tan}{\tg}
\def\dateline{\today}

\date{\dateline}

\def\W{{W}}
\begin{document}
\markboth{H. Epstein and U. Moschella}{Topological surprises in two-dimensional de Sitter QFT}

%
%

\title{Topological surprises in de Sitter QFT in two-dimensions.}

\author{Henri Epstein \\
{\small IHES. 35, Route de Chartres} \\
{\small Bures-sur-Yvette, 91441, France}\\
\\ Ugo Moschella
\\
{\small Disat, Universit\`a dell'Insubria, Como,  Italy}\\
}

\maketitle


\begin{abstract}
Motivated by the study of soluble models of quantum field theory, we illustrate 
a new type of topological effect by comparing the constructions of canonical Klein-Gordon quantum fields on the 
two-dimensional de Sitter spacetime as opposed to its double covering.
We show that the while commutators of the two fields coincide locally, the global topological differences make the theories drastically different. 
Many of the  well-known features of de Sitter quantum field theory disappear. In particular there is nothing like a Bunch-Davies vacuum.
Correspondingly, even though the local horizon structure is the same for the two universes there is no Hawking - Gibbons thermal state. 
Finally, there is no complementary series of fields.
\end{abstract}



\section{Introduction}

In spite of a large amount of papers published on the subject, 
quantum field theory on the de Sitter universe  is still in its infancy and needs to be  better understood. 
This study is so more urgent and necessary as it becomes more evident that the role played 
by this geometry in cosmology and astroparticle physics is at least comparable to the one played by Minkowski space in the usual relativistic quantum theory of fields.  This state of affaires might in a way recall the beginning of scientific cosmology \cite{desitter} where the de Sitter universe ({\em model B}) was confronted to  the Minkowski spacetime ({\em model  C}) and, of course, to the Einstein universe ({\em model A}) in the debate that gave rise to scientific cosmology. Now the challenge is to understand dS QFT  and  the cosmological constant from a quantum perspective.

Very often the study of quantum theories on curved spacetimes is confined to the construction (explicit when possible) of the two point correlation functions and to the consequences which can be derived thereof, such as for instance the study of the corresponding renormalized energy-momentum tensor. In the de Sitter case, attempts at perturbative renormalization also exist based either on the Schwinger-Keldysh formalism or on a  Euclidean approach and the role of the infrared divergences and their resummation is debated in the literature\cite{Akhme,0,1,2,3,4,5,6,7,8,9,10,11,12,13,14,15,16,17,18,19,20,21,22}. 
However the situation is still not as clear as one might desire.  
It would perhaps be useful to proceed beyond perturbation theory and explore - as much as possible - 
non-perturbative QFT in the presence of curvature. In particular, interesting new  phenomena might be related to global topological features and have no counterpart at all in flat spacetime; in that respect Minkowski QFT provides little or no clue to uncover them.

There are several non-perturbative roads to QFT. One of them is the study of soluble models in two spacetime dimensions, the best known being the Schwinger \cite{sch} and the Thirring\cite{thir,kla} models. In contrast with the inextricable complexity of any model of interacting fields in four dimensions, these models are sufficiently simple to allow for their explicit solution. Although this possibility is based on special properties of the dimension being two, which do not hold true any more if $d>2$,  these models have been theoretical laboratories which have made possible the discovery of phenomena and mechanisms that have subsequently revealed themselves to be essential characteristics of realistic models in four dimensions.
For example the Schwinger model, which corresponds to two-dimensional quantum electrodynamics, has allowed for the pre-discovery of some of the most important phenomena expected from quantum chromodynamics such as asymptotic freedom and confinement.

We have recently started  to investigate the same models formulated on the two-dimensional de Sitter universe  ${dS_2}$ with the aim to establish which one of their properties (if any) survive in the presence of curvature, starting from integrability itself. This study  may also throw a new light on some of the difficulties encountered in perturbation theory. 
In the first very preliminary step we have considered the (massless or massive) free Dirac  fields, and the results were  already  surprising to us\cite{Epstein1}. The point is that the two-dimensional de Sitter manifold is homotopically nontrivial and admits  two inequivalent spin structures. Correspondingly, there are two possible choices of Dirac fields which may be either periodic (Ramond) or anti-periodic (Neveu-Schwarz) w.r.t spatial rotations of an angle $2\pi$. When quantizing those fields, a requirement of de Sitter covariance (in a certain generalized sense)  may be implemented only in the anti-periodic case \cite{Epstein1}. A consequence of this fact is that the Thirring-de Sitter model admits covariant solutions\cite{Epstein2} only for the Neveu-Schwarz antiperiodicity condition.   The double covering of the de Sitter manifold $\widetilde{dS_2}$ naturally enters in the arena of QFT through that door.

The manifold $\widetilde{dS_2}$ is in itself a complete globally hyperbolic manifold. It carries a natural action of $SL(2,R)$, the double covering of  $SO_0(1,2)$, the pseudo-orthogonal group that acts on ${dS_2}$. The Lorentzian geometry of $\widetilde{dS_2}$ is locally indistinguishable from that of ${dS_2}$ but the global properties are quite different. This fact has profound consequences  at the quantum level. We present a few of them in this paper  by considering the simplest possible model of  quantum field theory, namely a free massive Klein-Gordon field.

\section{Geometrical setup}
The two-dimensional de Sitter group $G=SO_0(1,2)$ is the component  connected to the identity of 
the (Lorentz) pseudo-orthogonal group acting on  the three-dimensional Minkowski  spacetime $M_3$ with metric $\eta_{\mu\nu}$ = diag$(1,-1,-1)$.   The Iwasawa decomposition $KNA$ of a generic element is parametrized as follows: 
\begin{eqnarray}
&& g =k(\ttheta) n(\lambda)  a(u)= \cr  &&
\begin{array}{lll} \left(
 \begin{array}{ccc}
 1 & 0 & 0 \\
 0 & \cos \ttheta & \sin \ttheta \\
 0 & -\sin \ttheta & \cos \ttheta \\
\end{array}
\right)  \left(
\begin{array}{ccc}
1+ \frac{\lambda ^2}{2}  & \frac{\lambda  ^2}{2} & -\lambda  \\
 -\frac{\lambda ^2}{2} & 1-\frac{\lambda ^2}{2} & \lambda  \\
 -\lambda  & -\lambda  & 1 \\
\end{array}
\right)
 \left(
\begin{array}{ccc}
 \cosh u & -\sinh u & 0 \\
 -\sinh u & \cosh u & 0 \\
 0 & 0 & 1 \\
\end{array}
\right). &
\end{array} \cr && 
\end{eqnarray}
This factorization provides coordinates $(\lambda, \ttheta) $ to points of the coset space $G/A$  which is  topologically a cylinder. 
 $G$ acts on $G/A$ by left multiplication 
$
g\, :  \, (\lambda,\ttheta) \to (\lambda',\ttheta').
$
The  simple geometrical interpretation of all the above transformation rules 
may be seen by using the standard representation of the two dimensional de Sitter spacetime as an hyperboloid 
$
dS_{2}=\left\{ X \in {M}_{3}:\ (X^{0})^{2}-(X^1)^{2}-(X^2)^{2}=- 1 \right\}. \label{ds}
$
The parametrization of $dS_2$ corresponding to $(\lambda, \ttheta)$
\begin{equation}
X(\lambda,\ttheta)= \left\{
\begin{array}{l}
X^0= -\lambda , \\
X^1= \lambda  \cos \ttheta+\sin \ttheta ,\\
X^2= \cos \ttheta-\lambda  \sin \ttheta .\\
\end{array} 
\right.\label{pariwa}
\end{equation}
The coordinate system $(\lambda,\ttheta)$ is not orthogonal 
\begin{equation}
{\d}s^2 = \left.\left({\d X^0}^2-{\d X^1}^2-{\d X^2}^2\right)\right|_{dS_2} = -2 \d\lambda \d\ttheta  -\left(\lambda ^2+1\right)\d\ttheta^2.
\end{equation} 
The left action of $SO_0(1,2)$ on the cosets coincides with the linear action of  $SO_0(1,2)$ restricted to the manifold $dS_2$ and  the coset space $G/A$ is actually identical to the two-dimensional de Sitter manifold.

A similar construction shows that the covering manifold $\widetilde{dS_2}$ is a quotient space of
 the double covering of $SO_0(1,2)$ i.e. the spin group 
\begin{equation}
Sp(1,2)\cong SL(2,R)=
\bigl\{g\in SL(2,C) :\ \ g= 
\left(
\begin{array}{cc}
 a & i \,b \\
 i \,c & d \\
\end{array}
\right), \ \ \ a,b,c,d\in R,\ \ \ a d+bc=1\bigr \} .
\label{Sp}
\end{equation}
$Sp(1,2)$ and  $SL(2,R)$ are conjugate subgroups of $SL(2,C)$.
Events of  $dS_2$ are  transformed  by similarity: 
$
\slashed X' =  g\slashed X g^{-1}
\label{Sp2}
$
where as usual  
$
\slashed X= \gamma^\alpha X_\alpha 
$
and $\{\gamma^\mu,\gamma^\nu\} = 2 \eta^{\mu \nu}$.
The covering projection
$
g\rightarrow {\Lambda(g)^\alpha}_\beta = {1\over 2}{\rm tr}(\gamma^{\alpha}g\gamma_{\beta}g^{-1})
\label{lorentzmap}
$
of $Sp(1,2)$ onto $SO_0(1,2)$ is coherent 
with the above action in the sense that 
$
g\slashed X g^{-1} =  \cancel{\Lambda(g) X}. 
$

Let us now, as before, write the Iwasawa decomposition of $Sp(1,2)$:
\begin{equation}
 g=k(\zeta)\,n(\lambda)\,a(\chi) = \left(
\begin{array}{cc}
\cos\frac \zeta 2 & i \sin\frac \zeta 2 \\
i \sin\frac \zeta 2  & \cos\frac \zeta 2 \\
\end{array}
\right) \left(
\begin{array}{cc}
 1 & i \lambda\\
 0 & 1 \\
\end{array}
\right) \left(
\begin{array}{cc}
e^{\frac \chi 2}& 0 \\
 0 &e^{-\frac \chi 2}\\
\end{array}
\right); \label{iwadec}
\end{equation}
the parameters $\zeta,\lambda$ and $\chi$ are related to $a,b,c$ and $d$ by easily calculable relations; 
here $0\leq\zeta<4 \pi$ and $\lambda$ and $\chi$ are real and unrestricted. 
The above decomposition provides a natural parametrization $\tilde X (\lambda, \zeta)$ of the points of the symmetric space $Sp(1,2)/A $ .

The group $Sp(1,2)$ acts on the coset space by left multiplication: $
g \, \tilde X (\lambda, \zeta)\rightarrow \tilde X (\lambda', \zeta') $.
The Maureer-Cartan form 
 gives to  $Sp(1,2)/A $ 
 a natural left invariant Lorentzian metric that looks identical to the previous one  
\begin{equation}
\d s^2 = -2 \d\lambda \d\zeta  -\left(\lambda ^2+1\right)\d\zeta^2 \label{dscov}
\end{equation}
with the only difference that now $0\leq \zeta< 4\pi$. Finally the map
$p : \tilde X (\lambda, \zeta) \rightarrow X(\lambda,\zeta)$ is a covering map. 
The symmetric space $Sp(1,2)/A $  can be thus  identified with the double covering $\widetilde{dS_2}$ of the two-dimensional de Sitter universe; the spin group  $Sp(1,2)$ now acts faithfully on  $\wt{dS_2}$ by left multiplication and can be interpreted   as an isometry group of spacetime transformations i.e. a relativity group.

\section{Canonical quantization}

The first task in quantizing a (boson) field on a globally hyperbolic manifold ${\cal M}$ is  the construction of the commutator
$
[\phi(x),\phi(x')] = C(x,x'),
$
a bivariate distribution on  ${\cal M}$ which
has to vanish coherently with the notion of local causality inherent
to ${\cal M}$:  $C(x,x')=0$ for "spacelike separated"
points. In the case of a Klein-Gordon field 
the commutator is the {\em unique} 
solution of the  KG equation with initial conditions 
set by the the equal-time canonical commutation relations (CCR's):
\begin{eqnarray} && \left[\Box_{x} +m^2\right] C_{\lambda}(x,x') = 0 \ \  \left[\Box_{x'} +m^2 \right] C_{\lambda}(x,x')=0 \\&& C_{\lambda}(t, {\x}, t', {\x'})|_{t = t'} = 0, \ \  {\partial \over \partial t'}\, C_{\lambda}(t, {\x}, t', {\x'})|_{t = t'} = i \ch t' \ \delta_\Sigma({\x}-{\x'}).\label{c.1}\end{eqnarray}
The quantum commutation relations have an algebraic
character: in particular,  the commutator is a c-number and does not depend on anything but the mass $m$ and the Lorentzian structure of the 
manifold. 

The recipes of canonical quantization \cite{Birrell} allow to construct the commutator in terms of a complete set of complex
classical solutions $u_i(x)$  of the KG equation (the modes - labeled here by an index $i$) 
normalized as follows:
\begin{equation}
(u_i,u_j) = \delta_{ij},\;\;\;\;( u^*_i,  u^*_j) = -\delta_{ij},
\;\;\;\;(  u_i, u^*_j) = 0, \label{kgbase}
\end{equation}
where $(\cdot , \cdot )$ denotes the Peierls symplectic form also called the KG inner product\cite{Birrell}.
The commutator may be expanded in terms of the above modes as follows
\begin{equation}
C(x,x') = \sum [u_i(x) u^*_i(x')-u_i(x')
u^*_i(x)] 
\end{equation}
and this expression does not depend on the particular choice of modes.
In the following we will use the standard global spherical coordinate system 
\bea
X(t,\theta) = \left\{\begin{array}l
X^0=  \sinh t \\
X^1 =    \cosh t \sin \theta \\
X^2 =   \cosh t  \cos \theta 
\end{array}\right. .\ \ \ \ 
 \label{coorspheric}
\eea
The relation between  the angles $\zeta$ and $\theta$ is 
$ \zeta= \arctan \sinh t +\theta.$ The dS metric now reads 
\begin{equation}
ds^2 =  dt^2- \cosh^2 t \, d\theta^2 .
\end{equation}
 The same metric considered for $\theta \in [0,4\pi]$ 
describes the geometry of $\wt{dS_2}$ (which of course is not embedded in $M_3$). 
Let us  write the KG equation in the above coordinates  
\begin{equation}
\Box \phi - \lambda(\lambda+1) \phi=
\frac{1}{\cosh t} \ \partial_t (\cosh t \ \partial_t \phi)
- \frac{1}{\cosh^2 t} \ \partial^2_{\theta} \phi - \lambda(\lambda+1)\phi =0 ,
\label{s.16}
\end{equation}
 introduce the complex variable $z= i \sinh t$
and  separate the variables by posing
$
\phi =  f(z) e^{{i l \theta}}:$
the mode $f(z)$  must solve the Legendre differential equation 
\beq(1-z^2) f''(z) -2zf'(z) + \lambda(\lambda+1) f(z) -{l^2 \over (1-z^2)}f(z) = 0.\label{s.19}\endq
The complex squared mass   
$m^2_\lambda = - \lambda(\lambda+1)$ is real and positive for  either  $  \lambda = -\frac {1}2 + i \nu $
or  $(\Im \lambda =0)\cap (-1<\Re \lambda <0 )  $.  
Two suitable  
independent solutions are the  Ferrers functions  $\P^{-l}_\lambda(z)$ and $\Q_\lambda^{-l}(z)$ (also called  
``Legendre functions on the cut'' \cite{bateman}) 
which  are holomorphic in the
cut-plane
$
\Delta_2 =
{\bf C}\setminus (-\infty-1] \cup [1,\infty) 
$
 and satisfy there the important  reality conditions
\begin{equation}
[\P_\lambda^{-l}(z) ]^*= \P_{\lambda^*}^{-l^*} ( z^*), \ \ \ [\Q_\lambda^{-l}(z) ]^*= \Q_{\lambda^*}^{-l^*} ( z^*).
\end{equation}
$\P_\lambda^{-l}(z)$  alone respects the symmetry  of the complex squared mass  $m^2_\lambda = - \lambda(\lambda+1)$:
$
\P_\lambda^{-l}(z) =\P_{-\lambda-1}^{-l}(z).\label{t.22.1}$
If $\lambda-l$ and $-(\lambda+l+1)$ are not non-negative integers,  
 $\P^{-l}_\lambda(z)$ and  $\P^{-l}_\lambda(-z)$
are also independent solutions of Eq. (\ref{s.19}).

Let us therefore focus on the (not normalized) modes 
\begin{eqnarray}
\phi_{ l}(t,\x) &= & 
[a_l\P^{- l }_{\lambda}(i\sh t) +b_l  \P^{- l }_{\lambda}(-i\sh t)]e^{{il \theta}} \cr
\phi^*_{ l}(t,\x)& =  &
[a^*_l\P^{- l }_{\lambda}(-i\sh t) +b^*_l  \P^{- l }_{\lambda}(i\sh t)]e^{{-il \theta}} 
\label{modesab01}
\end{eqnarray}
where either $\lambda = -1/2 +i\nu$ or $\lambda$ real. The  KG product is defined as usual:
\begin{equation}
(f,g)_{KG} = 
i \int_\Sigma (f ^*\partial_\mu g -g\, \partial_\mu f ^* )  d\Sigma^\mu(x) = i \int_{\Sigma_0}  (f ^*\partial_t g -g\, \partial_t f ^* )  d\theta.
\label{kgprod}
\end{equation}
In the  $dS_2$  case the integral is over $\Sigma_{t=0}$ i.e. the interval $[0,2\pi]$ and $l$ is integer. When we consider fields on the covering manifold $\widetilde{dS_2}$ the  integral is over the interval $[0,4\pi]$ and $2l$ is integer.

The first condition to be imposed on the modes  is the orthogonality $(\phi_{l}, \phi^*{}_{ l'})_{KG} =0$;
it amounts to  the following conditions on the coefficients:
\begin{eqnarray}
&&\makebox{for $l\in{\Bbb Z}$} \ \ \ \ \quad  \quad a_l  b_{-l}  -b_l  a_{-l}  =0 \ \ \quad  \quad  \makebox{(for both $dS_2$ and $\widetilde{dS_2}$) , } \cr 
&&\makebox{for $l\in\frac 12 +{\Bbb Z}$} \ \  \left\{\begin{array}{l}
a_l  a_{-l}  -b_l  b_{-l} = c_l\sin (\pi  \lambda ) \\  a_l  b_{-l}  -b_l  a_{-l}  =c_l \sin(\pi l)\end{array} \right. \  \  \ \ \quad \makebox{(only for  $\widetilde{dS_2}$). } \label{condo}
\end{eqnarray}
 The constants $c_l$ are unrestricted by the above conditions. 
Besides we also  have that 
\begin{equation}
(\phi^{}_{  l},\phi^{}_{ l' })_{KG}
 = \frac{1}{N_l} \delta_{ll'} 
\ \ \ \makebox{with} \ \
N_l =  \frac {\gamma_l }{2 k\pi (|a_{ l}|^2 - |b_{ l}|^2)}  \label{kkgg12}
\end{equation}
where  $k=1$ for $dS_2$, $k=2$ for $\widetilde{dS_2}$
\beq
\gamma_l = \frac 12 {\Gamma(l-\lambda) \Gamma(1 +\lambda + l ) }.
\eeq
As a function of $l $, $ \gamma_l $ is always positive for $\lambda = -\frac 12 +i \nu$. For real $\lambda$ in the interval $(-1 , 0)$ $ \gamma_l $ is positive on the integers and negative on the half-integers.  Normalization amounts to imposing $N_l=1$ i.e. 
\begin{equation}
|a_{ l}|^2 - |b_{ l}|^2=  \frac {\gamma_l }{2 k\pi}  \label{kkgg12bis}
\end{equation}
and this sets restrictions on the possible values of $c_l$. The following formula gives the commutator both for the dS manifold ($k=1$) and its double covering ($k=2$):
\begin{eqnarray} 
 C_k(t,\theta,t',\theta')&=&\sum_{kl \in {\Bbb Z}}  \frac {\gamma_l }{2 k\pi (|a_{ l}|^2 - |b_{ l}|^2)} [\phi_{  l}(t,\x)
\phi^*_{  l}(t',\x') - \phi_{  l}(t',\x') 
\phi^*_{  l}(t,\x)] = \label{ccrcov}
\cr &&
\cr
&=&  \frac {i }{ k\pi }\sum_{kl \in {\Bbb Z}}   {\gamma_l }\Im [\P^{- l }_{\lambda}(i\sh t) \P^{- l }_{\lambda}(-i\sh t')] \cos{(l \theta-l \theta')} 
\cr
&+& \frac {i }{ k\pi } \sum_{kl \in {\Bbb Z}}    \frac {\gamma_l (|a_l|^2+|b_l|^2) }{(|a_{ l}|^2 - |b_{ l}|^2)}\Re [\P^{- l }_{\lambda}(i\sh t) \P^{- l }_{\lambda}(-i\sh t')]\sin{(l \theta-l \theta')} 
\cr
&+&  \frac {i }{ k\pi }\sum_{kl \in {\Bbb Z}}  \frac {2 \gamma_l }{ k\pi (|a_{ l}|^2 - |b_{ l}|^2)}\Re [ a_l b_l^* \P^{- l }_{\lambda}(i\sh t) \P^{- l }_{\lambda}(i\sh t')] \sin{(l \theta-l \theta')} .
\cr &&
\end{eqnarray}
The commutator  must however be a c-number and not depend on the choice of the coefficients $a_l$ and $b_l$. 
This means that  the second and third line in Eq. (\ref{ccrcov}) must vanish. Actually,  the second and third line in Eq. (\ref{ccrcov})  are the only ones to  contribute to the equal time commutator 
\begin{eqnarray} 
C_k(0,\theta,0,\theta')=   \frac{i }{k \pi}\sum_{kl \in {\Bbb Z}}   \frac {\gamma_l |a_l+b_l|^2} { |a_{ l}|^2 - |b_{ l}|^2}
 [ \P^{- l }_{\lambda}(0)]^2 \sin{(l \theta-l \theta'). } 
\end{eqnarray}
By using Eq. (\ref{condo}) we get that 
\beq
{ \frac{|a_l+b_l|^2}{|a_{l}|^2-|b_{l}|^2 } }=
 {\cot  \left(\frac 12 \pi(l+  \lambda)\right ) }{\tan\left(\frac 12 \pi  (\lambda - l)\right) } {\frac{|a_{-l}+b_{-l} |^2}{|a_{-l}|^2-|b_{-l}|^2} }
\eeq 
(here we took once more into account the hypothesis $\lambda = -1/2 +i\nu$ or $\lambda$ real). 
Since  
\begin{equation}
\frac{\gamma_l  \P^{- l }_{\lambda}(0)^2}{\gamma_{-l}  \P^{ l }_{\lambda}(0)^2} = \tan \left(\frac{1}{2} \pi  (\lambda +l)\right) \cot \left(\frac{1}{2} \pi 
   (\lambda-l )\right) \end{equation}
the coefficients of $\sin [l (\theta - \theta')]$ and of $\sin[-l(\theta - \theta')]$ are equal and $C(0,\theta,0,\theta')$ vanishes. 
On the other hand the second and third line in Eq. (\ref{ccrcov}) do not contribute to the CCR's:
\begin{eqnarray} 
\partial_{t'}C_k(t,\theta,t,\theta')|_{t=t'=0}&=&  -\frac{ i}{k\pi}  \sum_{k l \in {\Bbb Z}}    {\gamma_l}
 [ \P^{- l }_{\lambda}(0){ \P'}^{- l }_{\lambda}(0)] \cos{(l \theta-l \theta')} +\cr 
 &+&  \frac{ i}{k\pi}   \sum_{k l \in {\Bbb Z}} \sum  \frac {\gamma_l (a_l b_l^*-a^*_l b_l)}{|a_{l}|^2-|b_{l}|^2 }
 [ \P^{- l }_{\lambda}(0){ \P'}^{- l }_{\lambda}(0)] \sin{(l \theta-l \theta')} =\cr 
&=&  i  \sum_{k l \in {\Bbb Z}}   \frac {1 }{2k \pi} \cos{(l \theta-l \theta')} = i  \delta(\theta-\theta') \label{ccr}
\end{eqnarray}
where we used again Eq. (\ref{condo}). 
We deduce that the second and third line in Eq. (\ref{ccrcov}) vanish identically since they solve the KG equation with zero initial conditons. The covariant commutators may be finally re-expressed as follows: 
\begin{eqnarray} 
&& C_k(t,\theta,t',\theta') =
\sum_{kl \in {\Bbb Z}} \frac {\gamma_l}{2k\pi } [\P^{- l }_{\lambda}(i\sh t) \P^{- l }_{\lambda}(-i\sh t') - \P^{- l }_{\lambda}(-i\sh t) \P^{- l }_{\lambda}(i\sh t')] \cos{(l \theta-l \theta')}\  \cr && =
\sum_{kl \in {\Bbb Z}} \frac {\gamma_l }{2k\pi } [\P^{- l }_{\lambda}(i\sh t) \P^{- l }_{\lambda}(-i\sh t') - \P^{- l }_{\lambda}(-i\sh t) \P^{- l }_{\lambda}(i\sh t')] \exp{(i l \theta-i l \theta')}.  \label{coeff2} 
\end{eqnarray}
The second step follows from the symmetry of the generic term of the series at the right hand side of Eq. (\ref{coeff2}) 
under the exchange $l\to -l$. The commutator is  an entire function of the 
$\lambda$ variable. Indeed the poles of the coefficients
$\gamma_l(\lambda)$ are exactly compensated by zeros of the time-dependent factors 
and a de Sitter invariant canonical commutator exists also for integer 
values of the variable $\lambda$. Hence (\ref{coeff2}) holds for
all complex $\lambda$. 

\begin{figure}[h]
\centerline{{\includegraphics[width=3.9cm]{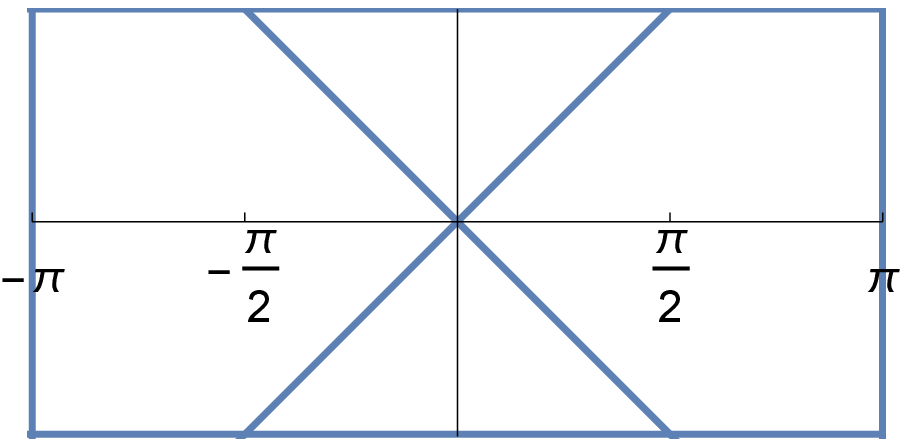}} \ \ \ \ \ \ \ \ \ {\includegraphics[width=7.8cm]{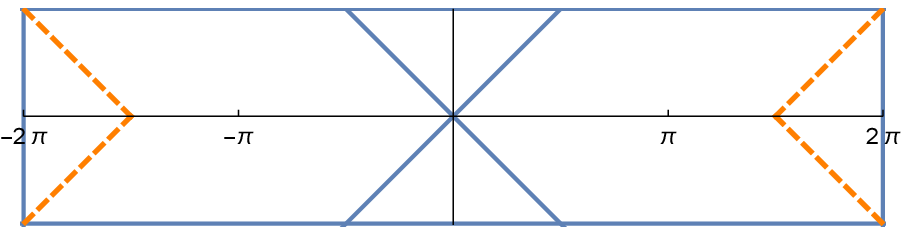}}}
\caption{Penrose diagrams for the de Sitter manifold and its double covering. If we extend the commutator $C_1$ to the covering manifold  by periodicity 
we see that $C_1$ and $C_2$ are non zero but coincide in the future and the past cones  of any given event (the origin).  $C_1$ and $C_2$  do not coincide when the second point explore the  second sheet. In particular $C_1$ does not vanish in those points that project in the future or the past cones  of the chosen origin (the region bordered by the dashed lines) while of course $C_2$ vanishes there, i.e. it is local in the sense of the covering manifold. 
\label{f1}}
\end{figure}

An interesting non trivial feature of the above expression is that the commutators $C_1$ and $C_2$ coincide when they are both non vanishing. This means that the field algebra of the de Sitter Klein-Gordon field and the one of the field on its covering are locally the same  differ globally (see figure \ref{f1}). The construction also guarantees  a priori that $C_1$ is $SO_0(1,2)$-invariant and that $C_2$ is $SL(2,R)$-invariant. A direct proof is also possible and instructive but it is not completely straightforward and will not be reproduced here.  

\section{Remarks about the construction of $SL(2,R)$-invariant states }
Once given the commutator, the crucial step to get a physical model is to represent 
the field $\phi$ as an operator-valued distribution in a
Hilbert space $\cal H$.  This can be done 
by finding a positive-semidefinite
bivariate distribution ${W}(x,y)$  solving the KG equation and the functional (non-differential) equation
\begin{equation}
{C}(x,x') = {\W}(x,x')- {\W}(x',x). \label{CR}
\end{equation}
Actually, $C$ and $W$ are not functions but distributions so the above equation must be understood in the sense of distributions.  
Given  a solution  $W$, the Gelfand-Naimark-Segal (GNS) procedure provides the
Fock space of the theory and a  representation of the field as a local operator-valued distribution (the word local here refers to {\em local commutativity}).
But uniqueness fails 
and  there are infinitely many inequivalent solution of Eq. (\ref{CR}). 
One immediate solution constructed in terms of the system (\ref{kgbase})
is the "vacuum" of the modes $u_i$:
\begin{equation}
W(x,x') = \sum u_i(x)  u^*_i(x').  \label{kg0}
\end{equation}
The standard theory of Bogoliubov canonical 
transformations provides infinitely many other, possibly inequivalent,  vacua   by
specifying the corresponding two-point functions in terms of two operators $a$ and $b$ as follows: 
\begin{equation}
W_{a,b}(x,x') = \sum\, [a_{ij
}u_j(x) + b_{ij} \, u^*_j(x)][a^*_{il} u^*_l(x')
+ b^*_{il}\, u_l(x')] .\label{kgab}
\end{equation}
Since the commutator must not depend on the choice of $a$ and $b$, Eq. (\ref{CR}) 
tells us  the conditions $\sum (a_{ij}a^*_{il} - b^*_{ij} b_{il}) = \delta_{jl}$ and $
\sum(a_{ij}b^*_{il}- a_{il}b^*_{ij})=0.$ 

The states given by Eq.  (\ref{kgab}) are "pure states". By considering more general superpositions one may describe also mixed states\cite{ms,ms2} within the same formalism.

Let us now address the question of the existence of vacua that are invariant under $SL(2,R)$ by exploring the general class of rotation invariant pure states   naturally associated with the modes (\ref{modesab01}): 
\begin{eqnarray}
&&\twofcn (x,x') = 
\sum_{kl \in {\Bbb Z}} { [a_l\P^{- l }_{\lambda}(i\sh t) +b_l  \P^{- l }_{\lambda}(-i\sh t)]
[a^*_l\P^{- l }_{\lambda}(-i\sh t') +b^*_l  \P^{- l }_{\lambda}(i\sh t')] }e^{{il \theta-il \theta'}}. \cr && \label{tp0}  
\end{eqnarray}
$F(x,x')$ has the right commutator (a fortiori, it is local)  and is positive definite for masses parameter of principal series $\lambda= -\frac 12 + i \nu$ (we use this name for the mass  even when there is no group representation at all).
Invariance  $F(gx,gx') = F(x,x')$ for all $g\in SL(2,R)$ holds if and only if 
\beq
\delta F= \sin \theta \, \partial_t F +\tanh t\, \cos \theta\,  \partial_\theta F+\sin \theta' \, \partial_{t'} F +\tanh t'\, \cos \theta'\,  \partial_{\theta'} F = 0.  \label{conddd}
\eeq
This relation implies that the periodic and the anti periodic parts of $F(x,x')$  
\begin{eqnarray}
\twofcn^{(0)} (x,x') = \sum_{l \in {\Bbb Z}} N_l \phi_l(x) \phi_l^*(x') , \ \ \  \twofcn^{(\frac 12 )} (x,x') = \sum_{l \in {\frac 12+\Bbb Z}} N_l \phi_l(x) \phi_l^*(x').
\end{eqnarray}
have to be separately invariant.
After a considerable amount of work it is possible  to show that 
the  $SL(2,R)$-invariance condition  (\ref{conddd}) implies the following relations:
\begin{eqnarray}    |a_l|^2 }= c_1(\epsilon) \gamma_l,   \ \  {|b_l|^2 = c_2(\epsilon)  \gamma_l , \ \    a_lb_l^* = c_3(\epsilon)  \gamma_l e^{i\l\pi}\label{ccond} \end{eqnarray}
where $\epsilon = 0$ for $l\in{\Bbb Z}$ and  $\epsilon = 1$ for $l\in \frac 12+{\Bbb Z}$ (i.e. there are six independent constants).
The above equations, together with the normalization condition (\ref{kkgg12bis}), can be solved as follows:
\begin{eqnarray}
&& a_{ l} = \sqrt{ \frac {\gamma_l} {2\pi k}} \cosh \alpha_\epsilon, \ \ \ \  b_{ l} = \sqrt{ \frac {\gamma_l } {2\pi k}} \sinh \alpha_\epsilon e^{i\phi_\epsilon-il\pi}, \ \ \ \ \epsilon=0, 1. \label{alpha}
\end{eqnarray}
Here we took $a_l$ real without loss of generality.
\subsection{Pure de Sitter}
Let us examine whether the above equations are compatible with the requirement (\ref{condo}) imposed by local commutativity.
In the pure de Sitter case (as opposed to its covering)  $l$ is integer and the CCR's amount to:
\begin{eqnarray}
&& a_l  b_{-l}  -b_l  a_{-l}  =0  .
\end{eqnarray}
This condition imposes no further restriction and any choice of $\alpha_0$ and $\phi_0$ gives rise to a de Sitter invariant state which has the right commutator (relatively to the de Sitter manifold). These states are well-known: they are the so-called alpha vacua \cite{spindel,allen,mottola,tach}. 

Among them, there is a particularly important state corresponding to the choice $\alpha_0= 0$: this is the so-called Bunch-Davies vacuum\cite{Bunch,Chernikov,hawking,bm,bgm,tach} 
\begin{eqnarray}
W_{BD}(x,x')=  F^{(0)}_{\alpha_0=0}(x,x')= \sum_{l \in {\Bbb Z}}\frac { \gamma_l }{2\pi} \P^{- l }_{\lambda}(i\sh t) 
\P^{- l }_{\lambda}(-i\sh t') e^{{il \theta-il \theta'}} = \cr =
{\Gamma(-\lambda)\Gamma(\lambda+1)\over 4\pi}
\,P_{\lambda }(\zeta),
 \label{tbd}  
 \end{eqnarray}
where $P_\lambda(\zeta) $ is the associated Legendre function of the first kind \cite{bateman} and the  de Sitter invariant variable $\zeta$ is the scalar product $\zeta = X(t-i\epsilon,\theta) \cdot X'(t'+i\epsilon,\theta)$ in the ambient space sense.
Actually, $W_{BD}(x,x')$ admits an extension to the complex de Sitter manifold and satisfies there  the {\em maximal analyticity property} \cite{bm,bgm}:
it is holomorphic for all $\zeta\in \bC\setminus (-\infty,\ -1]$ i.e. everywhere except on the locality cut. 
This crucial property singles the Bunch-Davies vacuum out of all the other invariant vacua and has a very well known thermal interpretation\cite{hawking,bm}: the restriction of the Bunch-Davies state a wedge-like region is a thermal state at temperature $T=1/2\pi$. A similar property is expected in interacting theories based on an analogue of the Bisognano-Wichmann theorem\cite{bem}.

\subsection{Covering}
In the antiperiodic case the CCR's 
\begin{eqnarray}
&&\makebox{for $l\in\frac 12 +{\Bbb Z}$} \ \  \left\{\begin{array}{l}
a_l  a_{-l}  -b_l  b_{-l} = c_l\sin (\pi  \lambda ) \\  a_l  b_{-l}  -b_l  a_{-l}  =c_l \sin(\pi l)\end{array} \right. \end{eqnarray}
imply the following  relation between the constants $\alpha$ and $ \phi$ and the mass parameter $\lambda$ of the field:  
\beq
e^{i \phi } \sin (\pi  l) (-i \sinh (2 \alpha ) \sin (\pi  \lambda )-i \cosh (2 \alpha )
   \sin \phi +\cos \phi )=0. \label{condoo}
\end{equation}
 For $\lambda = -1/2+ i\nu$ there is only one possible solution  given by   
\begin{equation}
\coth 2 \alpha = \cosh \pi \nu, \ \ \ \phi = \frac \pi 2.
\end{equation}
We denote the corresponding two-point function $F^{(\frac 12) }_{\nu}(x,x')$. 
Note that the value $\alpha =0$,  that would correspond to the above-mentioned maximal analyticity property, is excluded: it would be attained only for an infinite value of the mass. 
On the other hand  Equation (\ref{condoo}) has no solution at all when  $\lambda$ is real: there is no invariant vacuum of the complementary series.

In conclusion, for $\lambda = -\frac 12 + i \nu$  the most general  invariant vacuum state is the superposition of an arbitrary alpha vacuuum (the periodic part) plus a fixed antiperiodic part  $F^{(\frac 12) }_{\nu}(x,x')$ as follows 
\begin{equation}
F(x,x') = F^{(0)}_{\alpha_0, \phi_0}(x,x') +  F^{(\frac 12) }_{\nu}(x,x').
\end{equation}
For $\lambda = -\frac 12 + i\nu$ there is no $SL(2,R)$ local invariant vacuum state.
\section{Concluding remarks}
One important fact to be stressed again is that the above states do not share the global analyticity property characteristic of the Bunch-Davis maximally analytic vacuum\cite{Birrell,bgm,bm,bem}. Correspondingly, restriction of the above states to the wedge  between the horizons does not give rise to a thermal state as it happens in the pure de Sitter case (only for the BD vacuum)\cite{hawking,bm,bgm,bem}. The de Sitter temperature has disappeared. 
It remains to explore  the most general class of mixed invariant states. Our preliminary results tell that this fact remains true: $SL(2,R)$ invariance plus locality are in conflict with the spectral condition (even at microlocal level).  
This example teaches something also for quantum fields in spacetime dimension higher than two. 
First, a topological defect would be enough to destroy global  properties such as the de Sitter temperature, and second a global topological feature  
of the spacetime could be revealed by a local quantum measurement.

\section*{Acknowledgments}
UM thanks the IHES where this work has been done.


\end{document}